\documentclass[11pt,twoside]{article}
\usepackage{asp2010}

\resetcounters

\aspvoltitle{Astronomical Data Analysis Software and Systems XXI} 
\aspvolauthor{Pascal Ballester and Daniel Egret}

\begin{document}

\title{Why don't we already have an Integrated Framework for the Publication and Preservation of all Data Products?}
\author{Alberto Accomazzi$^1$, S\'ebastien Derriere$^2$, Chris Biemesderfer$^3$ and Norman Gray$^4$}
\affil{$^1$Harvard-Smithsonian Center for Astrophysics, 60 Garden Street,
Cambridge, MA 02138, USA}
\affil{$^2$Observatoire astronomique de Strasbourg
11, rue de l'Universit\'e, 67000 Strasbourg France}
\affil{$^3$American Astronomical Society, 2000 Florida Ave., NW
Suite 400, Washington, DC 20009-1231, USA}
\affil{$^4$Astronomy Group, School of Physics and Astronomy,
University of Glasgow, Glasgow, G12 8QQ, UK}

\begin{abstract}
Astronomy has long had a working network of archives supporting the curation of publications and data.  
The discipline has already created many of the features which perplex other areas of science:
(1) data repositories: (supra)national institutes, dedicated to large projects; a culture of user-contributed data; practical experience of long-term data preservation;
(2) dataset identifiers: the community has already piloted experiments, knows what can undermine these efforts, and is participating in the development of next-generation standards;
(3) citation of datasets in papers: the community has an innovative and expanding infrastructure for the curation of data and bibliographic resources, and through them a community of authors and editors familiar with such electronic publication efforts; as well, it has experimented with next-generation web standards (e.g. the Semantic Web);
(4) publisher buy-in: publishers in this area have been willing to innovate within the constraints of 
their commercial imperatives.
What can possibly be missing?  Why don't we have an integrated framework for the publication 
and preservation of all data products already?  Are there technical barriers?  We don't believe so.  
Are there cultural or commercial forces inhibiting this?  We aren't aware of any.  
This Birds of a Feather session (BoF) attempted to identify existing barriers to the creation of such a framework, and attempted to identify the parties or groups which can contribute to the creation of a VO-powered data-publishing framework.
\end{abstract}

\section{Introduction}
This BoF session provided a forum for data providers, publishers, librarians and scientists to explore the issues surrounding the preservation, identification and citation of data products in astronomy.  These are small but critical steps towards the ultimate goal of identifying the right practices, resources, and incentives to ensure that the entire research lifecycle in astronomy is properly captured and described in the coming era of data-intensive astronomical research.  
The authors of this paper (and BoF panelists) were chosen to represent varied perspectives from different corners of our community, including: the ADS, the primary literature portal in astronomy; the CDS, one of the largest data curation hubs in the Virtual Obsevatory (VO); the AAS, the largest publisher of scholarly literature in astronomy; and the University of Glasgow, an institution involved in data management and resource discovery for the VO.
The following sections highlight some of the topics covered in the BoF. 

\section{The Missing Data}
Analysis of the research literature published over the past two decades in the major astronomical journals uncovers several examples of researchers putting together websites that describe and give access to the datasets they have collected or created under their institute/home pages.  While this shows a commendable willingness on the part of these scientists to share their work and further disseminate their research, we have seen too many examples of such URLs disappear or go in disrepair as people move and institutes reorganize their websites.  In some cases we have seen entire domains, which were originally set up to host scientific papers and related data products, taken over by squatters.  Surely we can do better than this. 

But the first question to answer is: why are scientists doing this?  We believe that there is a desire for researchers to be able to package and present their work in a way that they feel is appropriate.  Give the high level of technical savvy of astronomers, there's never been a short-term barrier to putting up web pages, nor has anyone in the community perceived that as being tricky, or worth reward.  This has probably made it difficult to persuade people that there {\emph is} an unmet curatorial challenge that we need to confront, and a preservation need which is not currently matched by our existing infrastructure.

Even when data is stored in what one would consider authoritative,
trusted archives, there is currently no guarantee that their location
will persist in the long run.  Most of the effort on the part of the
big projects and science centers that host their observations is in
reducing and serving the data to enable new discoveries.  But shifting
technologies, economic realities and organizational changes often
force resources to be moved or, even worse, mothballed if they are not
considered to be essential in today’s research environment.  Thus, we
can only realistically take implicit promises of long-term data
archival as what they are: well-intentioned plans which are
contingent on a number of factors, some of which are out of our
control.  At the same time we should take steps to ensure that our
system of archiving, sharing and linking resources is as resilient as
it can be while we keep a realistic view of the technological and
economic environment supporting our research efforts.

\section{Preservation, Persistence and Versioning}
One of the first steps we should take in organizing our network of research data is to future-proof our nomenclature system by assigning persistent data identifiers (PDIDs) to products that we want to preserve but whose archival and curation are expected to change.  While this view is non-controversial and widely supported, coming to concrete decisions on how this should be implemented is not as straightforward.  There are several questions surrounding the technical and social aspects related to minting PDIDs:
\begin{itemize}
\item If we take the broadest view of preservation as an essential step in support of the repeatability of the research process, then we should archive and assign a unique PDID to each data product.  If the dataset is recreated as a result of an updated software pipeline or new calibration data, this should be considered a new version of the data product and be assigned a different PDID.  Thus under this scenario, an archive would need to “freeze” and uniquely identify all versions of data products it stores, a very costly and unlikely scenario for most archives.
\item A minimalistic approach to preservation is one in which only the data products described in scientific publications are persisted in a preservation environment and PDIDs are assigned to them.  In a small but growing number of journals covering social sciences and biosciences, deposit of datasets linked to publication to a community-supported archive is a required step in the publication process.  This ensures the creation of an archival dataset, the minting of a PDID, and the linking of the published article to the corresponding dataset.
\item A third option is one that attempts to bridge the two approaches above.  Preserving each possible version puts serious burdens on archives and is not necessarily a realistic model, but freezing a particular version when we know that it is being cited seems possible.  The issue then becomes: how do we know that data product X, downloaded from archive Y at time T should be frozen if the paper discussing it won’t appear in the literature for another two years?
\end{itemize}
Archive managers and curators of large datasets have repeatedly mentioned the practical difficulty of the first approach, and have pointed out that requests for older versions of data products are few and far in between.  On the other hand, derived datasets such as catalogs, montages and data mashups are updated and versioned on a regular basis (the best-known case being the Sloan Digital Sky Survey).   One additional issue that is often heard when considering minting PDIDs is what the right level of granularity should be for an identifier.  Should we assign a PDID to a mosaic of images, or individual images? To an observation dataset or the individual exposures?  To an aggregation of data products used in a study or to the hundreds of individual files in it?  

It is probably the case that there is no one-size-fits-all answer to the questions of granularity, versioning and preservation.  We expect that as data products are cited in the literature, the citation and preservation requirements will become clearer.  For the time being, as long as there is a mechanism which supports the citation of complete datasets, then it's deployable.  And as long as it can be subsequently refined when experience provides concrete demand, then we should not delay its use.  We also believe that we will greatly benefit from engaging the digital library community and leverage their experience when crafting a new set of data-centric citation and preservation policies.

\section{Citing and Linking Data}
What should data citations look like?  There are three models to be considered:
\begin{itemize}
\item Cite data as we cite articles: assign basic metadata to data products (author, title), a persistent identifier (DOI or other), and list them in the reference section, along with all other papers.  This is the model recommended for citing data in Dryad \citep{doi:10.1525/bio.2010.60.5.2} and DataVerse \citep{Crosas-Dlib}, among others.
\item Cite data as we cite websites: find out what their (hopefully persistent) URI is and mention it in your paper as an inline reference or a footnote.  This is the model that was adopted to cite data from the NASA archives via ADS \citep{2011fpca.conf..135A}. 
\item Have a “data references” section in a paper: similar to the bibliographic reference list, such a section would list in an unambiguous way (and using standard formatting) all the data products that were used in the study of a paper.  This option would make it easy for publishers, curators and aggregaters to identify data citations in a way similar to how we identify bibliographic citations today.
\end{itemize}
There is also no reason why we should not want to (semi-)automatically extract “data” citations from the fulltext of existing papers, similar to what NED and SIMBAD are doing in identifying objects mentioned in articles.  If we establish a consistent nomenclature for referring and linking to data products then this will be a much easier task.  
Hence it is essential that we as a community agree on well-defined standards for data identifiers and enable the creation of a corresponding registry to support this effort.  This will enable citations to data products to be recognized and identified in a way similar to what we currently do for citations to papers.

\section{Conclusions}
That there is a need for curated datasets, preserved indefinitely for scholarly purposes, is axiomatic by now. The details of a fully-fledged data preservation environment are difficult to discern at this point, but there are sensible places to begin, and from such places the mechanisms can evolve safely. The scientific and scholarly communities are vigorously pursuing possibilities, and there are many fruitful opportunities, such as this session at ADASS, for interested parties to engage in discussions about the promises and the challenges of broadly-scaled data preservation.

\bibliography{B1_v2}

\begin{thebibliography}{}
\expandafter\ifx\csname natexlab\endcsname\relax\def\natexlab#1{#1}\fi
\expandafter\ifx\csname url\endcsname\relax
  \def\url#1{\texttt{#1}}\fi
\expandafter\ifx\csname urlprefix\endcsname\relax\def\urlprefix{URL }\fi
\providecommand{\eprint}[2][]{\url{#2}}

\bibitem[{{Accomazzi}(2011)}]{2011fpca.conf..135A}
{Accomazzi}, A. 2011, in Future Professional Communication in Astronomy II,
  edited by {A.~Accomazzi}, 135. \eprint{arXiv:1103.4295}

\bibitem[{Crosas(2011)}]{Crosas-Dlib}
Crosas, M. 2011, D-Lib Magazine, 17

\bibitem[{Vision(2010)}]{doi:10.1525/bio.2010.60.5.2}
Vision, T.~J. 2010, BioScience, 60, 330

\end{thebibliography}

\end{document}